\pdfoutput=1

\documentclass[final,times,5p]{elsarticle}

\usepackage{graphicx}
\usepackage{amsmath,amssymb}
\usepackage{txfonts}
\usepackage{natbib}
\usepackage{hyperref}

\biboptions{square,sort&compress}

\journal{Chaos, Solitons \& Fractals}

\begin{document}

\begin{frontmatter}

\title{Equilibrium points and basins of convergence in the linear restricted \\
four-body problem with angular velocity}

\author[]{Euaggelos E. Zotos\corref{cor}}
\ead{evzotos@physics.auth.gr}

\cortext[cor]{Corresponding author}

\address{Department of Physics, School of Science, \\
Aristotle University of Thessaloniki, \\
GR-541 24, Thessaloniki, \\
Greece}

\begin{abstract}
The planar linear restricted four-body problem is used in order to determine the Newton-Raphson basins of convergence associated with the equilibrium points. The parametric variation of the position as well as of the stability of the libration points is monitored when the values of the mass parameter $b$ as well as of the angular velocity $\omega$ vary in predefined intervals. The regions on the configuration $(x,y)$ plane occupied by the basins of attraction are revealed using the multivariate version of the Newton-Raphson iterative scheme. The correlations between the attracting domains of the equilibrium points and the corresponding number of iterations needed for obtaining the desired accuracy are also illustrated. We perform a thorough and systematic numerical investigation by demonstrating how the parameters $b$ and $\omega$ influence the shape, the geometry and of course the fractality of the converging regions. Our numerical outcomes strongly indicate that these two parameters are indeed two of the most influential factors in this dynamical system.
\end{abstract}


\end{frontmatter}

\section{Introduction}
\label{intro}

There is no doubt that one of the most well investigated versions of the few-body problem is the circular or elliptic restricted (or not) three-body problem \cite{S67}. In the same vein, the planar restricted four-body problem describes the motion of a test particle with infinitesimal mass (with respect to the masses of the primaries) moving inside the gravitational field of three primary bodies. There are two main configurations regarding the position of the three primary bodies: (i) the Eulerian or linear configuration, where all three primaries lie on the same axis and (ii) the Lagrangian or triangular configuration, where the three primaries always lie at the vertices of an equilateral triangle. For the first configuration we have the case of the linear restricted four-body problem (LRFBP). Usually, for the corresponding configurations we use the term ``central configurations" due to the fact that the accelerations of the three primary bodies are proportional to the corresponding radius-vectors, while they are directed toward the common center of gravity \cite{M90}.

The four-body problem is a very important topic in celestial mechanics and dynamical astronomy for two main reasons. First of all, it is well known that about two thirds of the total stars in our Galaxy are in fact members of multi-stellar systems. Furthermore, approximately one fifth of these stars form triple systems, while roughly speaking about one fifth of these triples belong to more complex higher stellar systems (i.e., quadruple systems). The second reason is that the four-body problem allows us to describe and explain some of the most complicated dynamical phenomena that are encountered not only in our Solar System but also in exoplanetary systems. Therefore, studying the energetically allowed regions of motion, the stability as well as the families of periodic orbits in the four-body problem is fundamental in understanding the dynamical properties of multi-body stellar and exoplanetary systems \cite{SESS04}.

The LRFBP is also known as the Maranh\~{a}o-Llibre problem, as it was first studied in \cite{ML99}. In \cite{P07} the asymptotic solutions of the LRFBP have been studied, while the stable and the unstable manifolds around the hyperbolic Lyapunov periodic orbits which emanate from the equilibrium points have also been found. The effect of the radiation on the distribution of the families of periodic orbits, their stability, as well as the evolution of the families and their main features have been explored in \cite{KAE06,KAE07}, using the photogravitational version of the LRFBP. In the same vein, very recently in \cite{AAEP16} the position of the equilibria and their stability have been investigated in the linear restricted four-body problem where all three primary bodies were radiation emitters.

In dynamical systems knowing the basins of attraction associated with the equilibrium points is very important since this knowledge reveals some of the most inartistic properties of the system. For obtaining the basins of convergence we use an iterative scheme and we perform a scan of the configuration $(x,y)$ plane in order to determine from which of the equilibrium points (attractors) each initial condition is attracted by. The attracting domains in several types of dynamical systems have been numerically investigated. The Newton-Raphson iterative method was used in \cite{D10} to explore the basins of attraction in the Hill's problem with oblateness and radiation pressure, while in \cite{Z16} the multivariate version of the same iterative scheme has been used to unveil the basins of convergence in the restricted three-body problem with oblateness and radiation pressure. Furthermore, the Newton-Raphson converging domains for the photogravitational Copenhagen problem \cite{K08}, the electromagnetic Copenhagen problem \cite{KGK12}, the four-body problem \cite{BP11,KK14}, the ring problem of $N + 1$ bodies \cite{CK07,GKK09}, or even the restricted 2+2 body problem \cite{CK13} have been studied.

In this paper we shall work as in \cite{Z16}, thus following the same numerical techniques and methodology, and we will try to reveal the Newton-Raphson basins of attraction on the configuration $(x,y)$ plane for special case of the restricted four-problem where the three primary bodies are in linear configuration.

The present paper is organized as follows: In Section \ref{mod} we present the basic properties of the considered mathematical model. In section \ref{lgevol} the parametric evolution of the position and the stability of the equilibrium points is investigated with respect to the values of the mass parameter and the angular velocity. In the following Section, we conduct a thorough and systematic numerical exploration by revealing the Newton-Raphson basins of attraction of the LRFBP and how they are affected by the values of the mass parameter and the angular velocity. Our paper ends with Section \ref{disc}, where the discussion and the main conclusions of this work are presented.

\section{Properties of the mathematical model}
\label{mod}

\begin{figure}[!t]
\centering
\resizebox{\hsize}{!}{\includegraphics{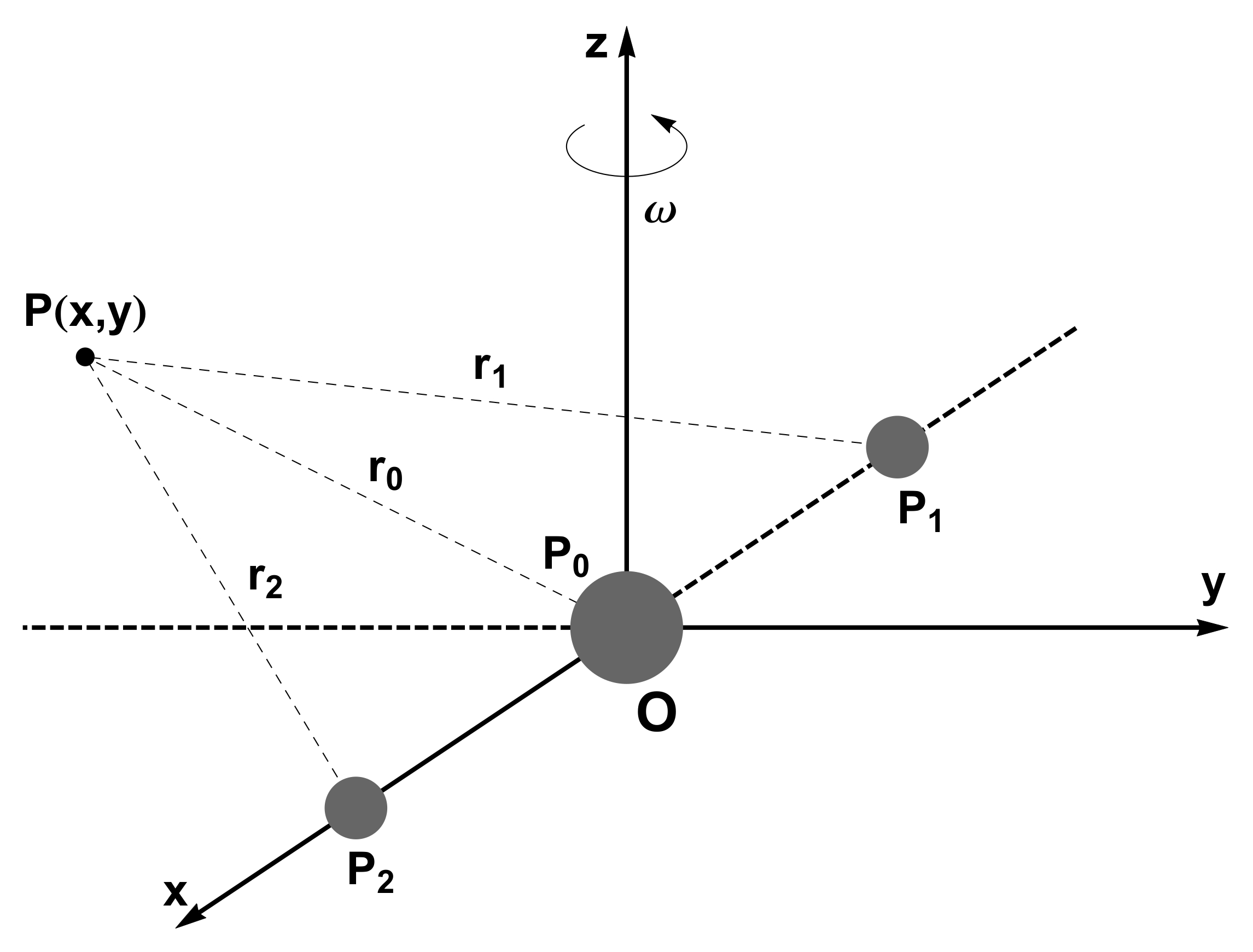}}
\caption{Linear configuration of the three primary bodies, moving in circular orbits around their common center of gravity.}
\label{lrfbp}
\end{figure}

We consider the case where a test particle $P$, with infinitesimal mass, moves under the gravitational filed of three primary bodies $P_0$, $P_1$, and $P_2$. It is assumed that the three primaries are in a collinear configuration on the $x$ axis. More precisely, primaries $P_1$ and $P_2$ have the same mass $m_1 = m_2 = m$ and they are located in symmetric positions with respect to the central primary body $P_0$, which has a different mass $m_0$, where $b = m_0/m$ is the so-called \textit{mass parameter}. The primary $P_0$ is placed between primaries $P_1$ and $P_2$ at the mass center $O$ of the system. It is interesting to note that for $b = 0$ (the central primary body is absent), the Copenhagen case of the restricted three-body problem is derived. The peripheral bodies $P_1$ and $P_2$ perform circular orbits around $P_0$ with constant angular velocity $\omega$ (see Fig. \ref{lrfbp}).

We adopt a planar synodic frame of reference having its origin at the central primary $P_0$. The $Ox$ axis joins the primaries directed towards $P_2$, while the $Oy$ axis complete the direct frame. We assume only planar motion of the test particle $P$ on the configuration $(x,y)$ plane.

The units of length, mass and time are chosen in such a way so that $\| P_1P_2 \| = 1$ and also $G m_1 = 1$. On this basis, the coordinates of the three primary bodies in the synodic frame of reference are: $P_0(0,0)$, $P_1(-x_p,0)$, and $P_2(x_p,0)$, where $x_p = 1/2$.

The effective potential function in a synodic system of coordinates is defined as
\begin{equation}
\Omega(x,y) = \frac{1}{\Delta} \left[ \frac{b}{r_0} + \sum\limits_{i=1}^2 \frac{1}{r_i} \right] + \frac{\omega^2}{2}\left(x^2 + y^2\right),
\label{pot}
\end{equation}
where
\begin{equation}
\Delta = 2\left(1 + 4b\right),
\end{equation}
while
\begin{align}
&r_0 = \sqrt{x^2 + y^2}, \nonumber\\
&r_1 = \sqrt{\left(x + x_p \right)^2 + y^2}, \nonumber\\
&r_2 = \sqrt{\left(x - x_p \right)^2 + y^2},
\label{dist}
\end{align}
are the distances to the respective primaries.

The equations describing the motion of an infinitesimal mass (test particle) read
\begin{align}
&\Omega_x = \frac{\partial \Omega}{\partial x} = \ddot{x} - 2\dot{y} = x \omega^2 - \frac{1}{\Delta} \left[\frac{b x}{r_0^3} + \frac{\left(x + x_p \right)}{r_1^3} + \frac{\left(x - x_p \right)}{r_2^3} \right], \nonumber\\
&\Omega_y = \frac{\partial \Omega}{\partial y} = \ddot{y} + 2\dot{x} = y \left[\omega^2 - \frac{1}{\Delta}\left(\frac{b}{r_0^3} + \frac{1}{r_1^3} + \frac{1}{r_2^3} \right) \right],
\label{eqmot}
\end{align}
where dots denote the time derivatives.

The equations of motion of the LRFBP have the following property: if $x = x(t)$ and $y = y(t)$ is a solution, then $x = x(-t)$ and $y = y(-t)$ is also a solution of the system. This is true because if we substitute $x \to - x$, $y \to - y$, $\dot{x} \to - \dot{x}$, $\dot{y} \to - \dot{y}$, $\ddot{x} \to - \ddot{x}$, and $\ddot{y} \to - \ddot{y}$ to the set of the differential equations (\ref{eqmot}) then the equations of motion remain unchangeable.

The system of the equations of motion admits the integral of the total orbital energy (also known as the Jacobi integral of motion)
\begin{equation}
J(x,y,\dot{x},\dot{y}) = 2\Omega(x,y) - \left(\dot{x}^2 + \dot{y}^2 \right) = C,
\label{ham}
\end{equation}
where $\dot{x}$ and $\dot{y}$ are the velocities, while $C$ is the Jacobi constant which is conserved.

\section{Parametric variation and stability of the equilibrium points}
\label{lgevol}

It is well known that the necessary and sufficient conditions for the existence of every equilibrium point are
\begin{equation}
\dot{x} = \dot{y} = \ddot{x} = \ddot{y} = 0.
\label{lps0}
\end{equation}
Therefore, the coordinates of the positions of all the coplanar equilibrium points of the LRFBP can be numerically derived by solving the following system of partial differential equations
\begin{equation}
\begin{cases}
\Omega_x(x,y) = 0 \\
\Omega_y(x,y) = 0
\end{cases}.
\label{lps}
\end{equation}

\begin{figure}[!t]
\centering
\resizebox{\hsize}{!}{\includegraphics{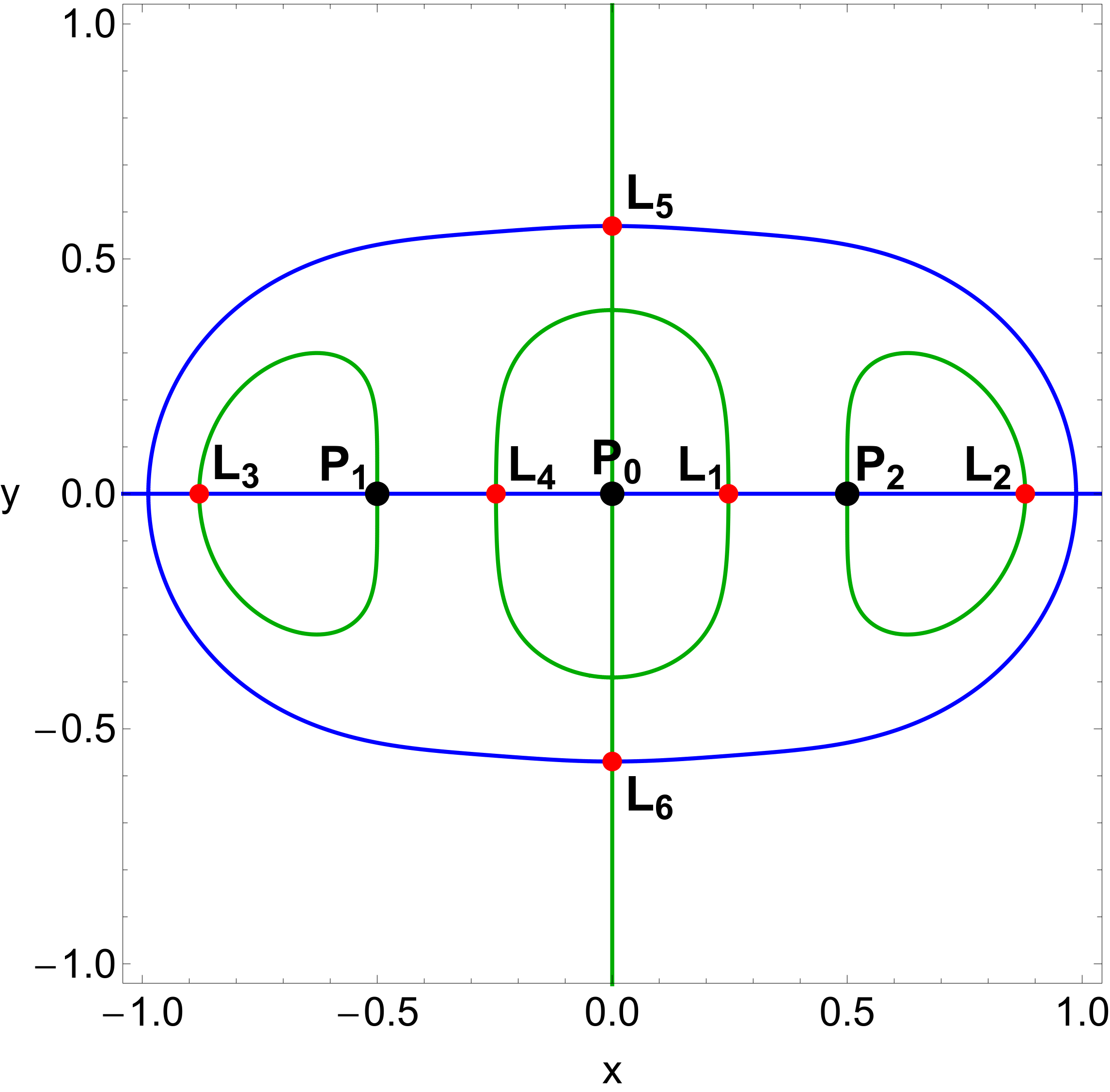}}
\caption{Locations of the positions (red dots) of the equilibrium points $(L_i, \ i = 1,6)$ through the intersections of $\Omega_x = 0$ (green) and $\Omega_y = 0$ (blue), when $b = \omega = 1$. The black dots denote the centers $(P_i, \ i=1,3)$ of the three primary bodies.}
\label{conts}
\end{figure}

As it is well know the classical restricted three-body problem has five equilibrium points. The LRFBP on the other hand, has six libration points. Four of them, $L_i$, $i = 1,...,4$ are collinear, while the other two $L_i$, $i = 5,6$ are triangular \cite{KAE07}. The first four equilibrium points lie on the primaries line. In particular, $L_1$ is on the positive $Ox$ axis between primaries $P_0$ and $P_2$, $L_2$ is on the same axis outside $P_2$, while the libration points $L_4$ and $L_3$ are symmetric to $L_1$ and $L_2$, respectively, with respect to the origin. Moreover, the equilibrium point $L_5$ is on the positive $Oy$ axis, while $L_6$ is on the negative vertical semi-axis. In Fig. \ref{conts} we see how the intersections of equations $\Omega_x = 0, \ \Omega_y = 0$ define, on the configuration $(x,y)$ plane, the positions of the six equilibrium points when $b = \omega = 1$.

\begin{figure*}[!t]
\centering
\resizebox{\hsize}{!}{\includegraphics{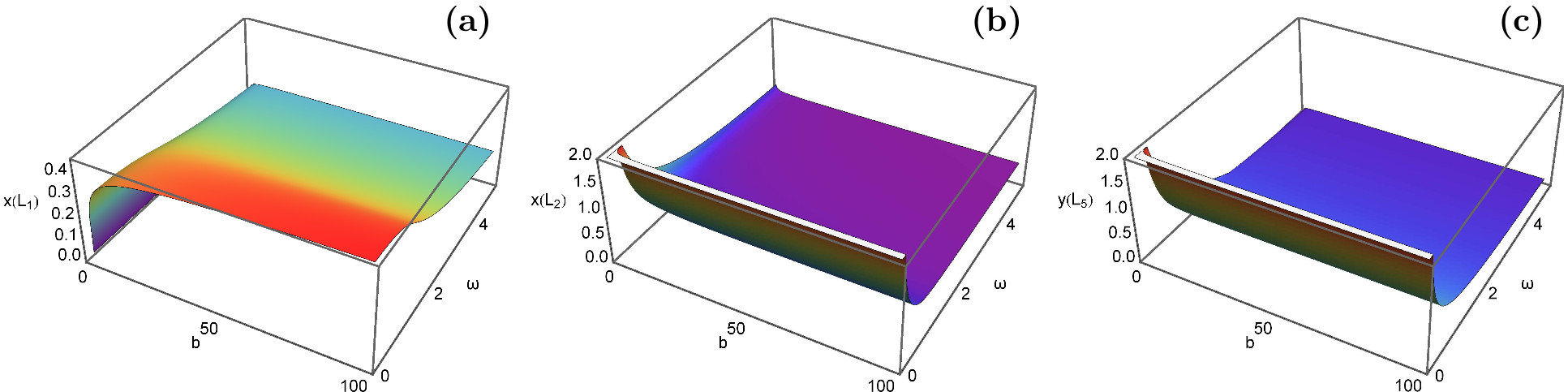}}
\caption{The space-evolution of the coordinates of the equilibrium points in the LRFBP when $b \in (0, 100]$ and $\omega \in (0, 5]$. Light reddish colors indicate high values of the coordinates, while dark blue/purple colors correspond to low values of the coordinates.}
\label{lgs}
\end{figure*}

In this investigation we shall reveal how the mass parameter $b$ and the angular velocity $\omega$ influence the positions of the equilibrium points, when they vary in the intervals $b \in (0, 100]$ and $\omega \in (0, 5]$. For this task we define a two-dimensional rectangular dense grid of $1024 \times 1024$ equally spaced initial conditions $(b_0,\omega_0)$. Then for every pair of initial conditions $(b_0,\omega_0)$ we numerically calculate the exact position of the equilibrium points. Our outcomes are illustrated in Fig. \ref{lgs}(a-c), where we present the space-evolution of the $x$ coordinates of $L_1$ and $L_2$ and of the $y$ coordinate of $L_5$. For the libration points $L_3$, $L_4$, and $L_6$ the results are entirely symmetric with respect to those of $L_2$, $L_1$, and $L_5$, respectively (only the sign of the coordinates changes).

In \cite{P07} it was shown that in the LRFBP with $\omega = 1$ all collinear points are unstable for every possible value of the mass parameter $b$, while the triangular points $L_5$ and $L_6$ are stable only when $b > b_{\rm crit} \simeq 11.72034889$. Now we will explore the stability of the equilibrium points when both parameters $b$ and $\omega$ vary in the above-mentioned intervals.

For determining the stability of an equilibrium point the origin of the frame of reference is transferred at its position $(x_0,y_0)$ following the transformation
\begin{align}
&x = x_0 + \xi, \nonumber\\
&y = y_0 + \eta.
\end{align}
Then we expand the system of equations of motion (\ref{eqmot}) into first-order terms with respect to $\xi$ and $\eta$ thus obtaining the linearized system which describes infinitesimal motions near an equilibrium point
\begin{equation}
\dot{{\bf{\Xi}}} = A {\bf{\Xi}}, \ \ {\bf{\Xi}} = \left(\xi, \eta, \dot{\xi}, \dot{\eta}\right)^{\rm T},
\label{ls}
\end{equation}
where ${\bf{\Xi}}$ is the state vector of the test particle with respect to the equilibrium points, while $A$ is the time-independent coefficient matrix of variations
\begin{equation}
A =
\begin{bmatrix}
    0 & 0 & 1 & 0 \\
    0 & 0 & 0 & 1 \\
    \Omega_{xx}^0 & \Omega_{xy}^0 & 0 & 2\omega \\
    \Omega_{yx}^0 & \Omega_{yy}^0 & -2\omega & 0
\end{bmatrix},
\end{equation}
where the superscript 0 at the partial derivatives of second order denotes evaluation at the position of the equilibrium point $(x_0, y_0)$.

The characteristic equation of the linear system (\ref{ls}) is quadratic with respect to $\Lambda = \lambda^2$ and is given by
\begin{equation}
\alpha \Lambda^2 + b \Lambda + c = 0,
\label{ceq}
\end{equation}
where
\begin{align}
&\alpha = 1, \nonumber\\
&b = 4\omega^2 - \Omega_{xx}^0 - \Omega_{yy}^0, \nonumber\\
&c = \Omega_{xx}^0 \Omega_{yy}^0 - \Omega_{xy}^0 \Omega_{yx}^0.
\end{align}

An equilibrium point is stable only when all roots of the characteristic equation for $\lambda$ are pure imaginary. Therefore the following three necessary and sufficient conditions must be simultaneously fulfilled
\begin{equation}
b > 0, \ \ c > 0, \ \ D = b^2 - 4 a c > 0,
\end{equation}
which ensure that the characteristic equation (\ref{ceq}) has two real negative roots $\Lambda_{1,2}$, which consequently means that there are four pure imaginary roots for $\lambda$.

\begin{figure}[!t]
\centering
\resizebox{\hsize}{!}{\includegraphics{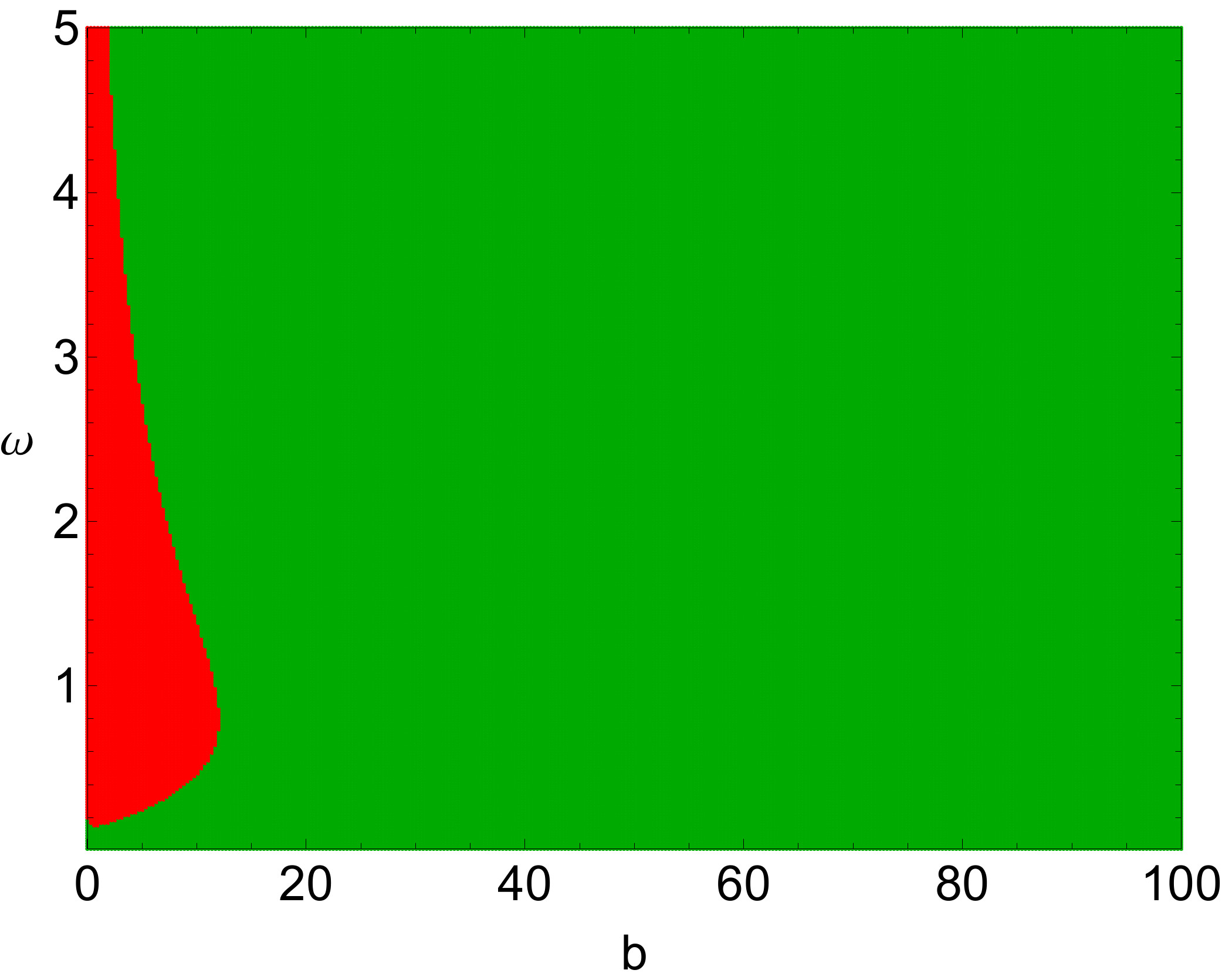}}
\caption{A color-coded grid on the $(b, \omega)$ plane illustrating the stability (green) or instability (red) of the triangular points $L_5$ and $L_6$ of the LRFBP.}
\label{clas}
\end{figure}

Knowing the exact position of the equilibrium points (see Fig. \ref{lgevol}) we can easily insert them into Eq. (\ref{ceq}), determine the nature of the fours roots and then derive the stability of the libration points. Our numerical calculations suggest that the four collinear points (for which $\Omega_{xy}^0 = \Omega_{yx}^0 = 0$) are unstable for all possible values of the mass parameter and the angular velocity. The triangular points on the other hand, can be either stable or unstable, depending of course on the particular values of $b$ and $\omega$. In Fig. \ref{clas} we present a two-dimensional color-coded grid on the $(b, \omega)$ plane thus revealing for which values of $b$ and $\omega$ the points $L_5$ and $L_6$ are stable or unstable. It is seen that the triangular points are mostly stable except for relatively low values of the mass parameter where they become unstable.

\section{The basins of attraction}
\label{bas}

There is no doubt that the most famous numerical method for solving systems of equations is the Newton-Raphson method. This method is also applicable to systems of multivariate functions $f({\bf{x}}) = 0$, through the iterative scheme
\begin{equation}
{\bf{x}}_{n+1} = {\bf{x}}_{n} - J^{-1}f({\bf{x}}_{n}),
\label{sch}
\end{equation}
where $J^{-1}$ is the inverse Jacobian matrix of the system of differential equations $f({\bf{x_n}})$, where in our case it is described in Eqs. (\ref{lps}).

With trivial matrix calculations (see e.g., the Appendix in \cite{Z16}) we can obtain the following iterative formulae for each coordinate
\begin{align}
&x_{n+1} = x_n - \left( \frac{\Omega_x \Omega_{yy} - \Omega_y \Omega_{xy}}{\Omega_{yy} \Omega_{xx} - \Omega_{xy} \Omega_{yx}} \right)_{(x_n,y_n)}, \nonumber\\
&y_{n+1} = y_n + \left( \frac{\Omega_x \Omega_{yx} - \Omega_y \Omega_{xx}}{\Omega_{yy} \Omega_{xx} - \Omega_{xy} \Omega_{yx}} \right)_{(x_n,y_n)},
\label{nrm}
\end{align}
where $x_n$, $y_n$ are the values of the $x$ and $y$ coordinates at the $n$-th step of the iterative process, while the subscripts denote the corresponding partial derivatives of first and second order of the effective potential function $\Omega(x,y)$.

\begin{figure}[!t]
\includegraphics[width=\hsize]{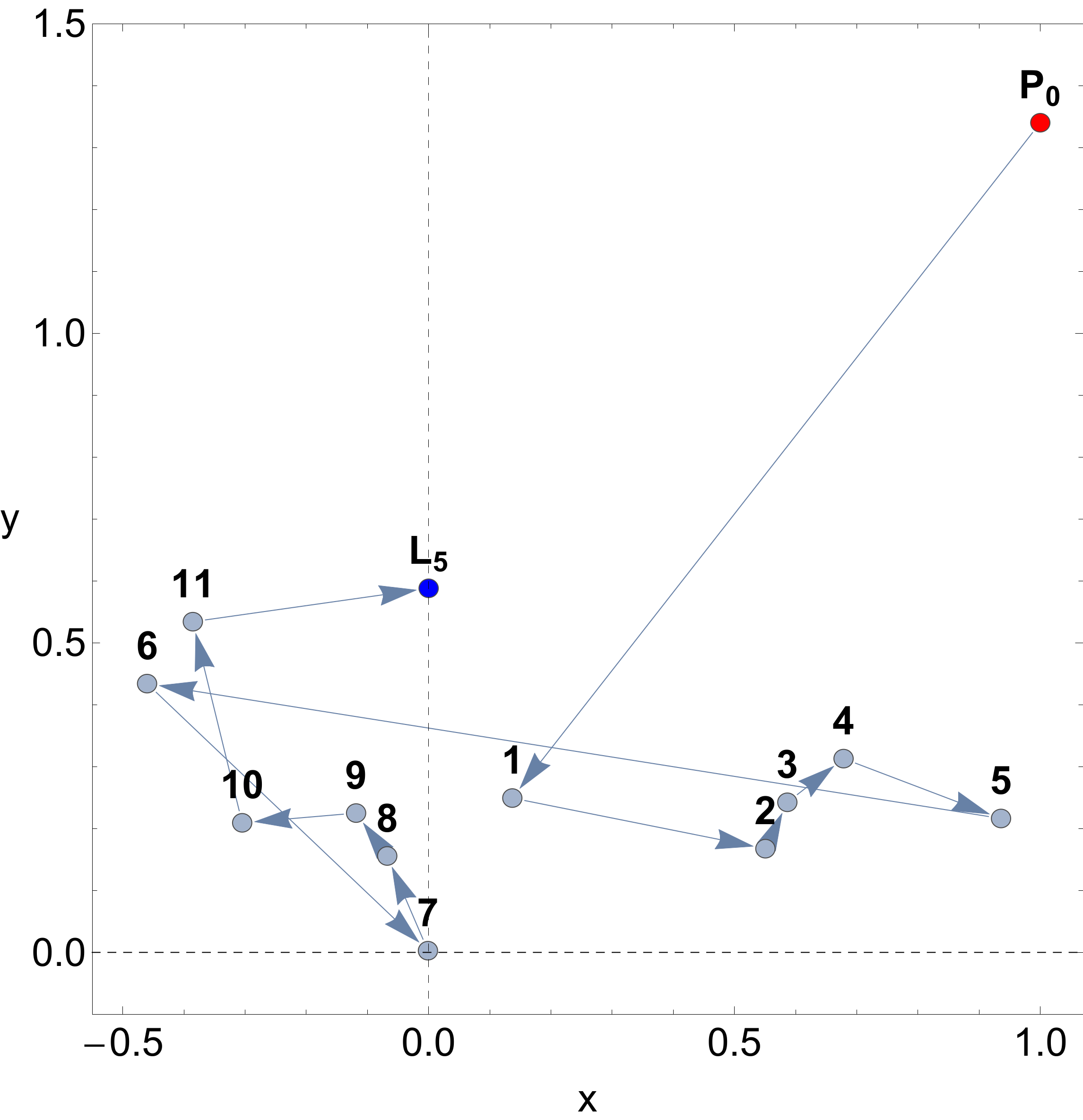}
\caption{A characteristic example of the consecutive steps that are followed by the Newton-Raphson iterator and the corresponding crooked path-line that leads to an equilibrium point $(L_5)$, when $b = \omega = 1$. The red dot indicates the starting point $P_0$ with $(x_0, y_0) = (1, 1.34)$, while the blue dot indicates the equilibrium point to which the method converged to. For this particular set of initial conditions the Newton-Raphson method converges after 12 iterations to $L_5$ with accuracy of eight decimal digits, while only three more iterations are required for obtaining the desired accuracy of $10^{-15}$.}
\label{nr}
\end{figure}

\begin{figure*}[!t]
\centering
\resizebox{\hsize}{!}{\includegraphics{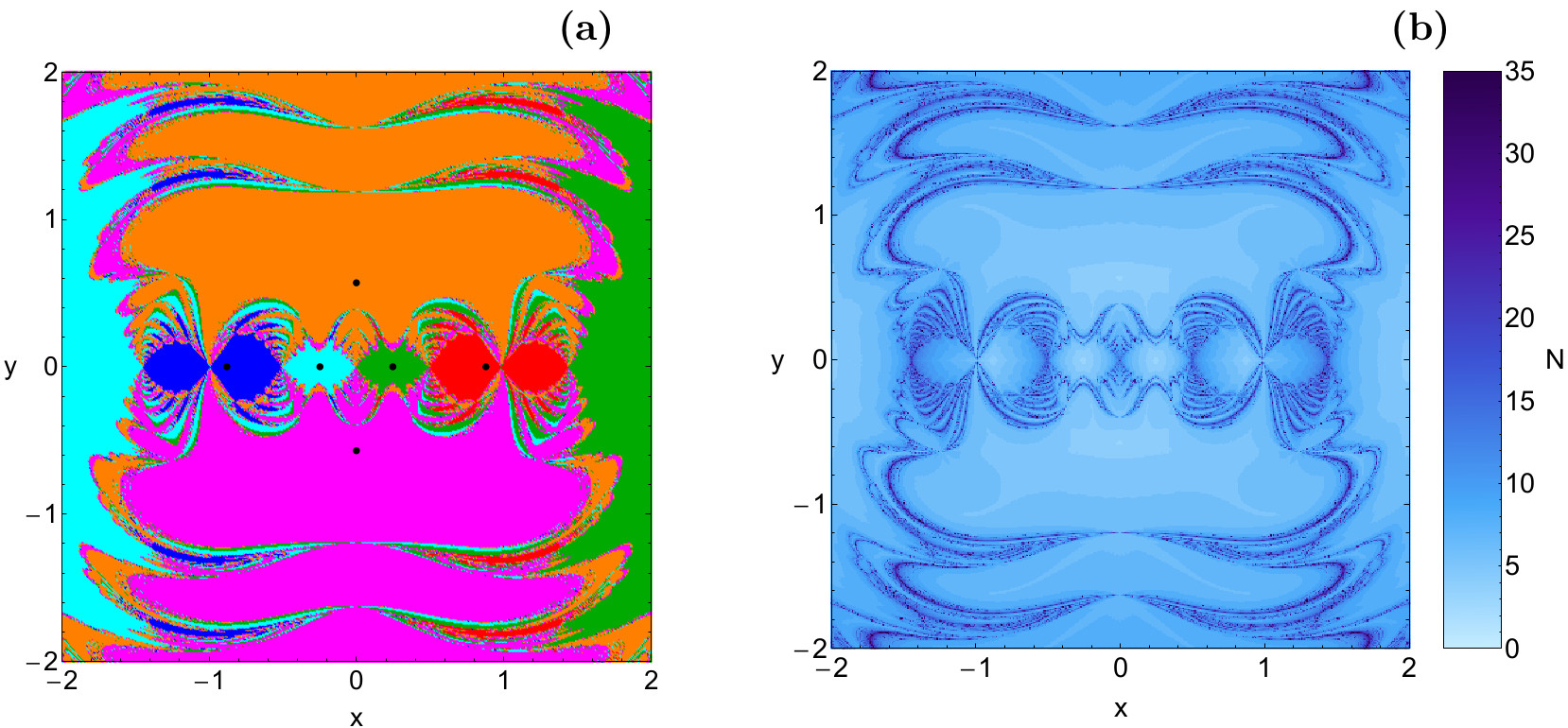}}
\caption{(a-left): The Newton-Raphson basins of attraction, on the configuration $(x,y)$ plane, for the LRFBP when $b = \omega = 1$. The positions of the six equilibrium points are indicated by black dots. The color code denoting the six attractors (equilibrium points) is as follows: $L_1$ (green); $L_2$ (red); $L_3$ (blue); $L_4$ (cyan); $L_5$ (orange); $L_6$ (magenta); non-converging points (white). (b-right): The distribution of the corresponding number $(N)$ of required iterations for obtaining the Newton-Raphson basins of attraction shown in panel (a).}
\label{sm}
\end{figure*}

The Newton-Raphson algorithm works as follows: an initial condition $(x_0,y_0)$ on the configuration plane activates the code and the iterative process continues until one of the equilibrium points of the system is reached, with some predefined accuracy. In most of the cases the successive approximation points create a crooked path line (see Fig. \ref{nr}). The initial condition may or may not converge to one of the libration points which act as attractors. If the crooked path leads to one of the equilibrium point then the iterative method converges for the particular initial condition. A Newton-Raphson basin of attraction\footnote{It should be clarified and clearly emphasized that the Newton-Raphson basins of convergence should not be mistaken, by no means, with the classical basins of attraction which exist in dissipative systems. The difference between the Newton-Raphson basins of convergence and the basins of attraction in dissipative systems is huge. This is true because the attraction in the first case is just a numerical artifact of the Newton-Raphson iterative method, while in dissipative systems the attraction is a real dynamical phenomenon, observed through the numerical integration of the initial conditions.} or convergence (also known as attracting region or domain) is composed of all the initial conditions that lead to a specific attractor (equilibrium point).

One may claim that knowing the basins of attraction of a dynamical system is an issue of paramount importance because these attracting regions may reflect some of the most important qualitative properties of the system in question. This can be justified by taking into account the fact that the derivatives of both first and second order of the effective potential function $\Omega(x,y)$ are included in the iterative formulae (\ref{nrm}).

For revealing the structures of the basins of attraction on the configuration $(x,y)$ plane we define a dense uniform grid\footnote{The initial conditions corresponding to the three centers $(P_0, P_1, P_2)$ of the primaries are excluded from the grid because for these values the distances $r_i$, $i = 0,1,2$ to the primaries are zero and therefore several terms of the formulae (\ref{nrm}) become singular.} of $1024 \times 1024$ initial conditions (nodes), which will be used as the initial values of the numerical algorithm. The iterative procedure begins and stops only when an accuracy of $10^{-15}$ regarding the position of the attractors has been achieved. A double scanning of the configuration plane is performed in order to classify all the available initial conditions that lead to a specific equilibrium point (or attractor). While classifying the initial conditions we also record the number $N$ of required iterations in order to obtain the aforementioned accuracy. It is evident that there is a strong correlation between the required number of iterations and the desired accuracy; the better the accuracy the higher the required iterations. In this study we set the maximum number of iterations $N_{\rm max}$ to be equal to 500.

\begin{figure*}[!t]
\centering
\resizebox{0.70\hsize}{!}{\includegraphics{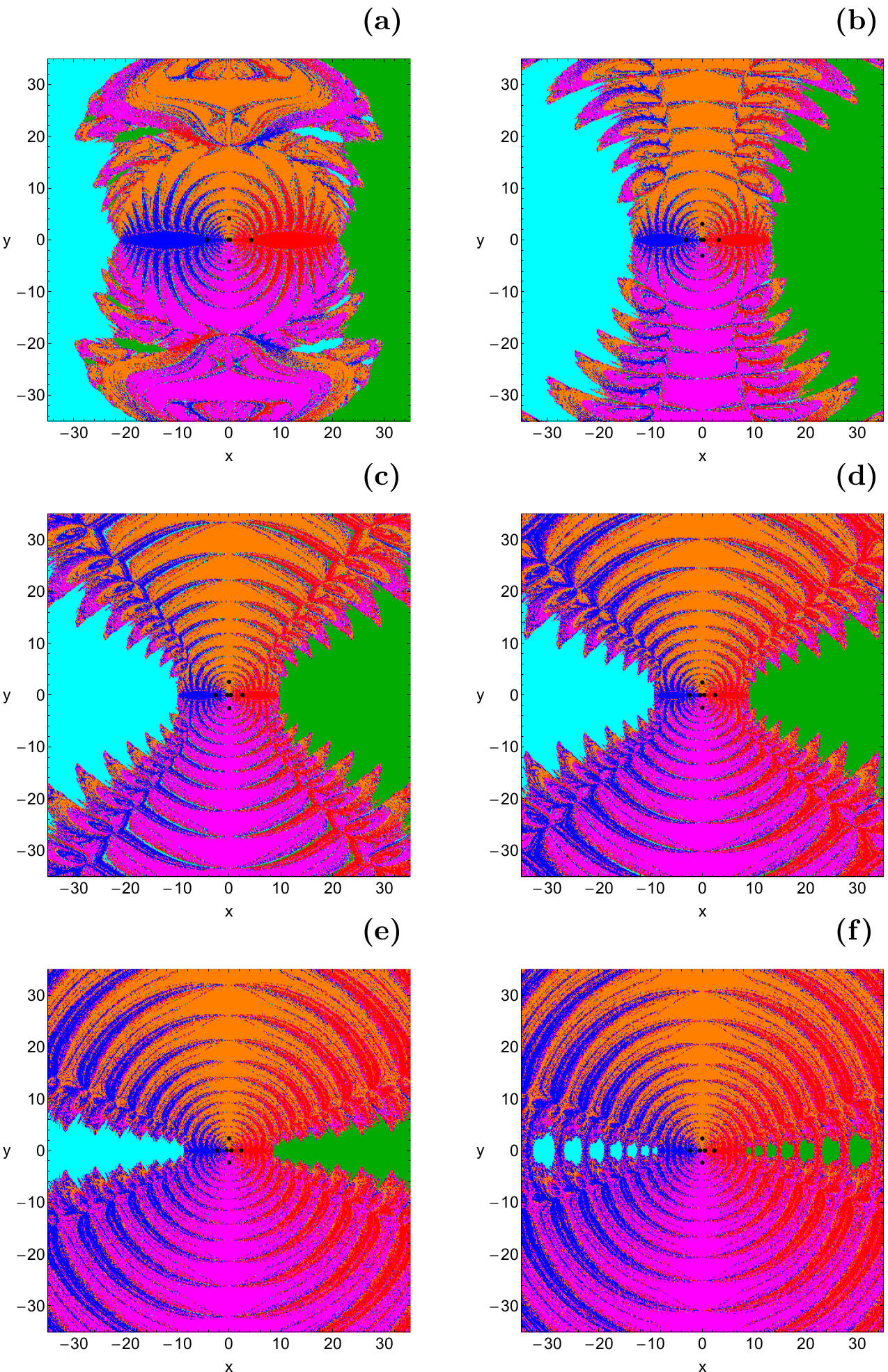}}
\caption{The Newton-Raphson basins of attraction on the configuration $(x,y)$ plane for the first case, where $\omega = 0.1$. (a): $b = 0.1$; (b): $b = 1$; (c): $b = 5$; (d): $b = 10$; (e): $b = 50$; (f): $b = 100$. The positions of the six equilibrium points are indicated by black dots. The color code, denoting the six attractors and the non-converging points, is as in Fig. \ref{sm}.}
\label{oml}
\end{figure*}

\begin{figure*}[!t]
\centering
\resizebox{0.85\hsize}{!}{\includegraphics{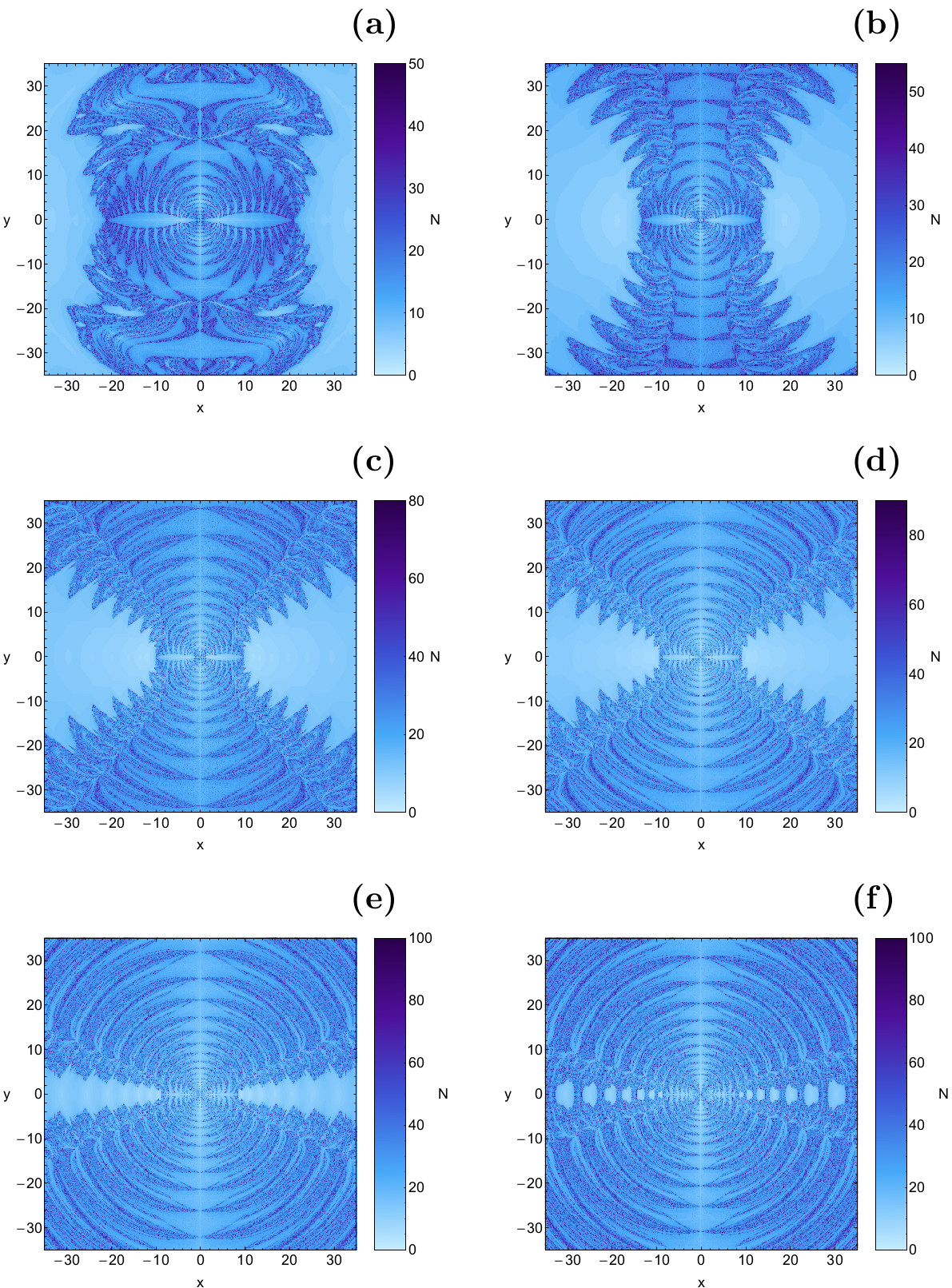}}
\caption{The distribution of the corresponding number $(N)$ of required iterations for obtaining the Newton-Raphson basins of attraction shown in Fig. \ref{oml}(a-f). The non-converging points are shown in white.}
\label{omln}
\end{figure*}

\begin{figure*}[!t]
\centering
\resizebox{0.70\hsize}{!}{\includegraphics{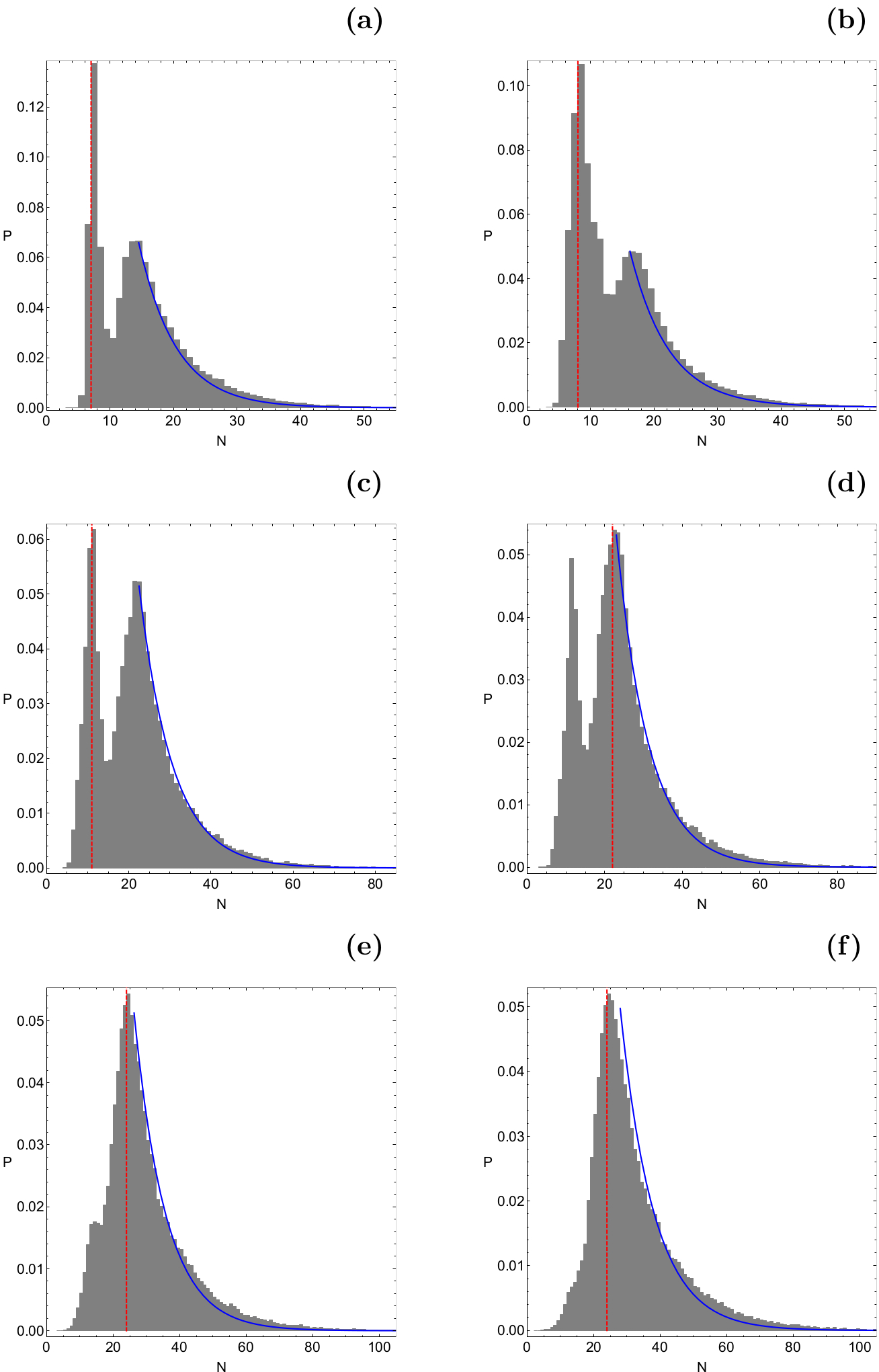}}
\caption{The corresponding probability distribution of required iterations for obtaining the Newton-Raphson basins of attraction shown in Fig. \ref{oml}(a-f). The vertical dashed red line indicates, in each case, the most probable number $(N^{*})$ of iterations. The blue line is the best fit for the right-hand side $(N > N^{*})$ of the histograms, using a Laplace probability distribution function.}
\label{omlp}
\end{figure*}

\begin{figure*}[!t]
\centering
\resizebox{\hsize}{!}{\includegraphics{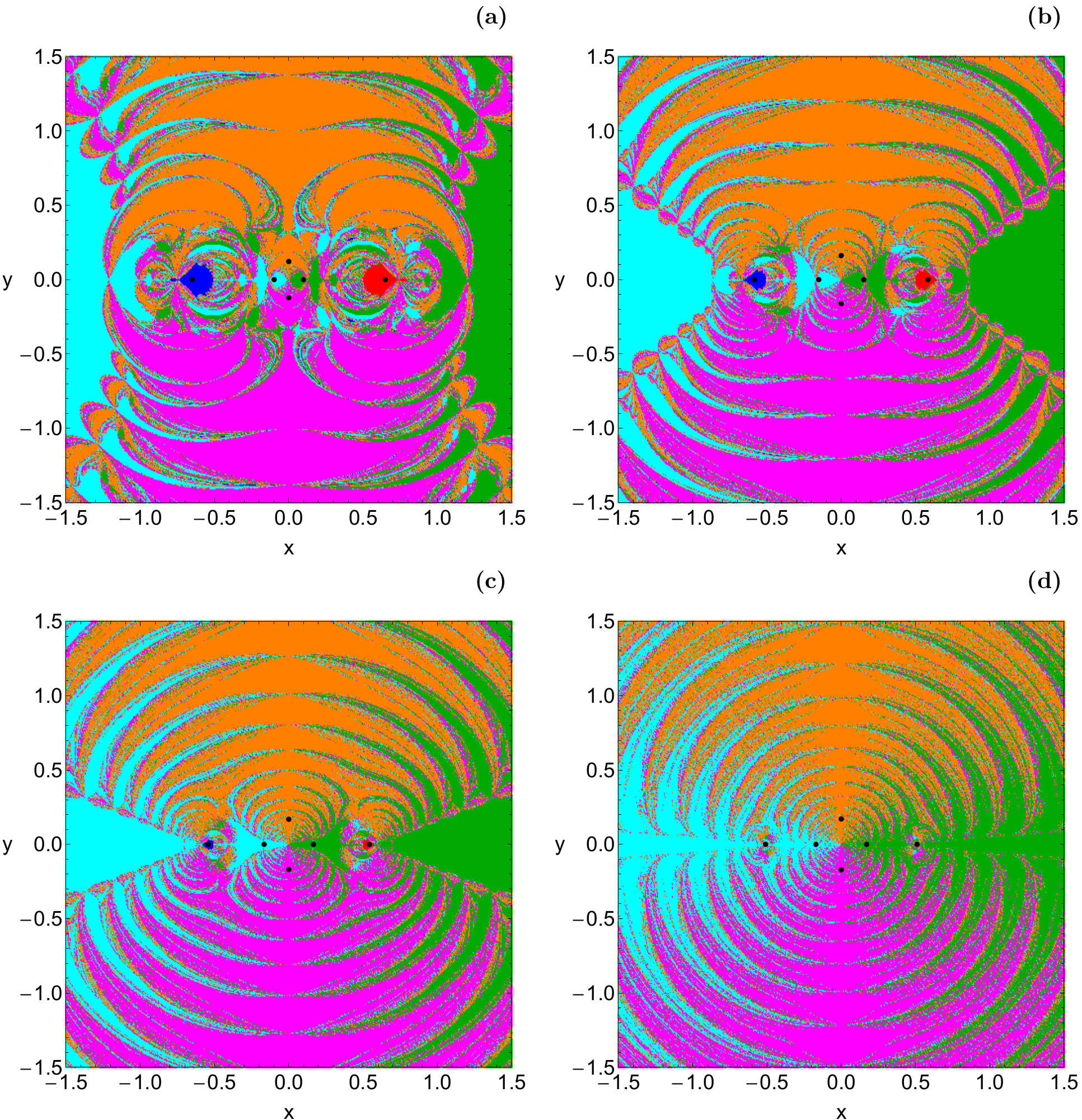}}
\caption{The Newton-Raphson basins of attraction on the configuration $(x,y)$ plane for the second case, where $\omega = 5$. (a): $b = 0.1$; (b): $b = 1$; (c): $b = 5$; (d): $b = 100$. The positions of the six equilibrium points are indicated by black dots. The color code, denoting the six attractors and the non-converging points, is as in Fig. \ref{sm}.}
\label{omh}
\end{figure*}

\begin{figure*}[!t]
\centering
\resizebox{\hsize}{!}{\includegraphics{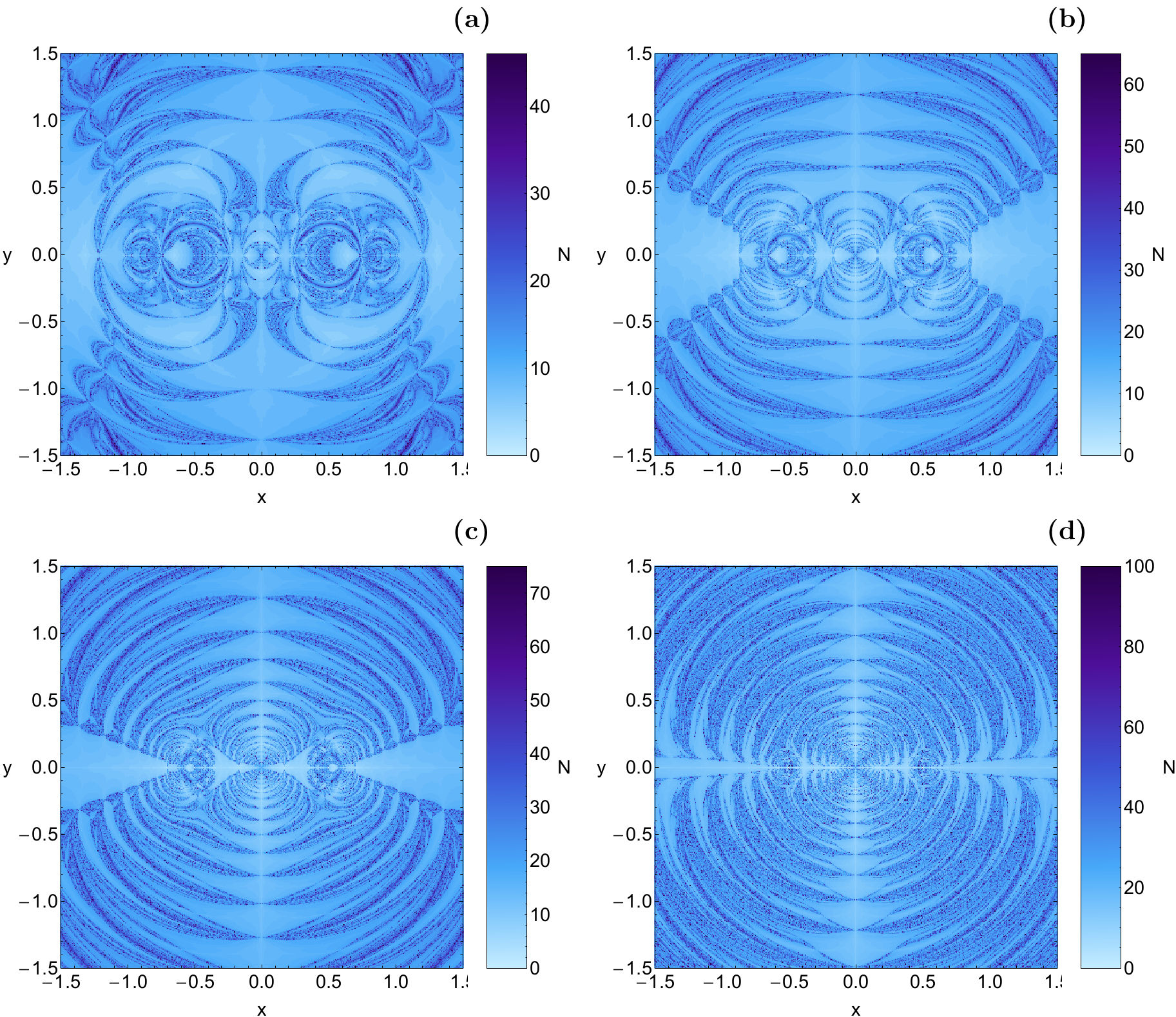}}
\caption{The distribution of the corresponding number $(N)$ of required iterations for obtaining the Newton-Raphson basins of attraction shown in Fig. \ref{omh}(a-d). The non-converging points are shown in white.}
\label{omhn}
\end{figure*}

\begin{figure*}[!t]
\centering
\resizebox{\hsize}{!}{\includegraphics{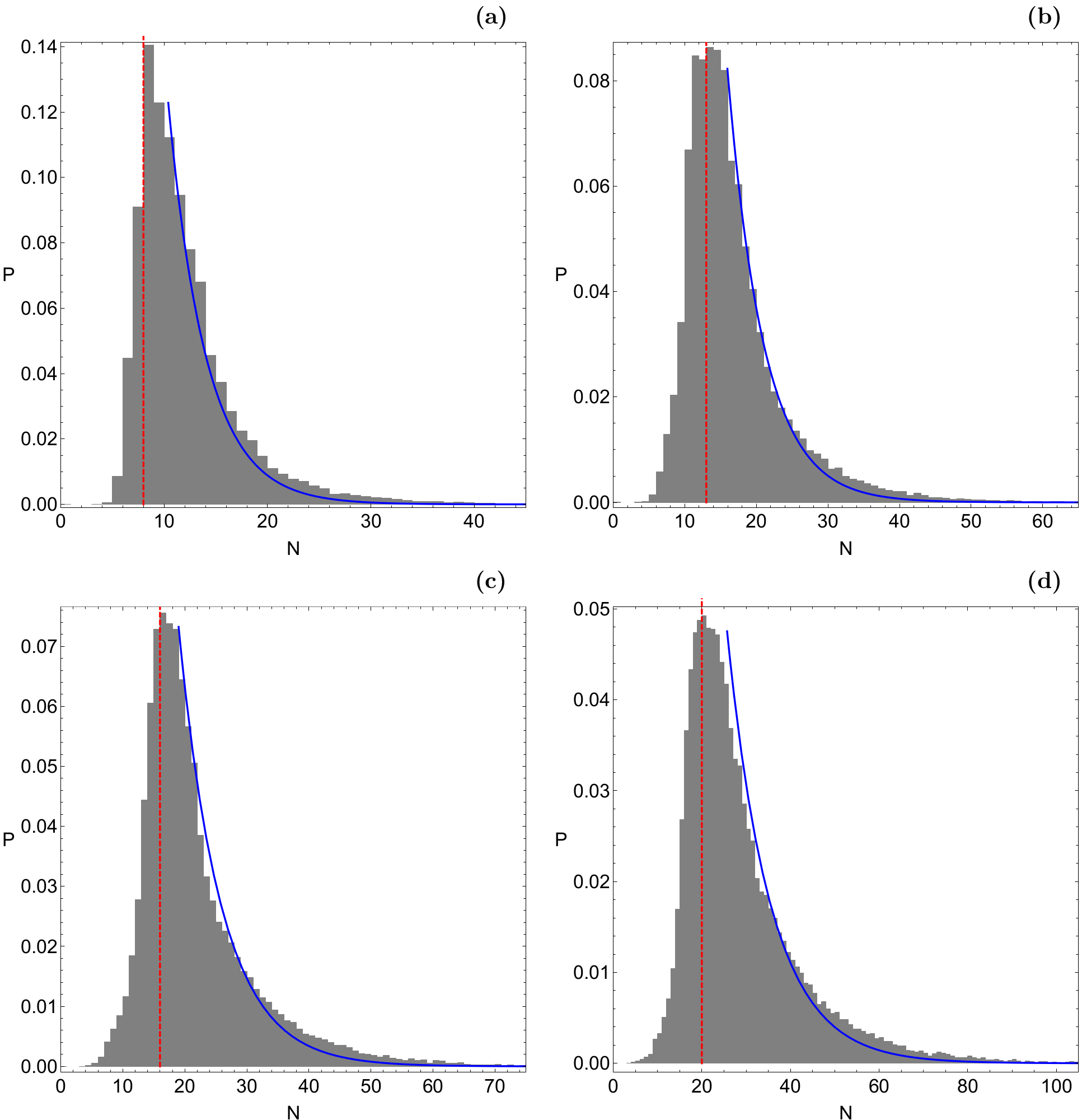}}
\caption{The corresponding probability distribution of required iterations for obtaining the Newton-Raphson basins of attraction shown in Fig. \ref{omh}(a-d). The vertical dashed red line indicates, in each case, the most probable number $(N^{*})$ of iterations. The blue line is the best fit for the right-hand side $(N > N^{*})$ of the histograms, using a Laplace probability distribution function.}
\label{omhp}
\end{figure*}

In panel (a) of Fig. \ref{sm} we present the Newton-Raphson basins of attraction when $b = \omega = 1$, which means that all three primaries have equal masses. For each basin of convergence we use different color, while the positions of all the attractors (equilibrium points) are pinpointed by small black dots. All non-converging points are shown in white. In panel (b) of the same figure the distribution of the corresponding number $(N)$ of iterations required for obtaining the desired accuracy is given using tones of blue. Looking the color-coded plot in Fig. \ref{sm}a we may say that the shape of the basins of convergence corresponding to equilibrium points $L_2$ and $L_4$ look like exotic bugs with many legs and many antennas, while the shape of the basins of attraction corresponding to all other libration points look like butterfly wings.

In the following we shall try to determine how the mass parameter $b$ as well as the angular velocity $\omega$ influence the structure of the Newton-Raphson basins of attraction, considering two cases regarding value of the angular velocity. For the classification of the initial conditions on the $(x,y)$ plane we will use modern color-coded diagrams. In these diagrams, each pixel is assigned a specific color according to the particular attractor (equilibrium point).

\subsection{Case I: Low angular velocity}
\label{ss1}

Our investigation begins with the case where the angular velocity has a relatively low value, that is when $\omega = 0.1$. In Fig. \ref{oml}(a-f) we present a collection of color-coded plots illustrating the Newton-Raphson basins of convergence for six values of the mass parameter when $b = \{0.1, 1, 5, 10, 50, 100\}$. It is seen that well-formed basins of convergence cover the majority of the configuration $(x,y)$ plane. However, the boundaries of all these basins exhibit a highly fractal\footnote{When we state that a domain displays fractal structure we simply mean that it has a fractal-like geometry however, without conducting any specific calculations for computing the fractal dimensions as in \cite{AVS09}.} structure and we may say that they behave as a ``chaotic sea". The meaning of chaos is justified taking into account that if we choose a starting point $(x_0,y_0)$ inside these fractal areas we will observe that the choice is highly sensitive. In particular, even a slight change in the initial conditions leads to a completely different final destination (different attractor). This implies that in these areas it is almost impossible to predict from which of the libration points each initial condition is attracted by.

Our computations suggest that in the LRFBP the basins of convergence of all six attractors extend to infinity. In the classical restricted three-body problem on the other hand, the only basins of attraction with infinite area are the attracting domains corresponding to the central equilibrium point $L_1$.

As the value of the mass parameter $b$ increases the following important phenomena take place in the configuration $(x,y)$ plane: (i) the structures of the attracting domains corresponding mainly to the triangular points $L_5$ and $L_6$ become wider and for $b \geq 100$ the two parts are joined together; (ii) the area of the bug-like structures of the basins of convergence corresponding to collinear points $L_2$ and $L_3$ is constantly reduced; (iii) the area of the attracting regions of libration points $L_2$ and $L_3$ which are present between the attracting domains of the triangular points $L_5$ and $L_6$ increases.

The distribution of the corresponding number $(N)$ of iterations required for obtaining the desired accuracy is provided in Fig. \ref{omln}(a-f), using tones of blue. It is more than evident that initial conditions inside the basins of attraction converge relatively fast $(N < 20)$, while the slowest converging points $(N > 50)$ are those in the vicinity of the basin boundaries. In the same vein, in Fig. \ref{omlp}(a-f) the corresponding probability distribution of iterations is presented. The definition of the probability $P$ is the following: if we assume that $N_0$ initial conditions $(x_0,y_0)$ converge to one of the available attractors, after $N$ iterations, then $P = N_0/N_t$, where $N_t$ is the total number of initial conditions in every color-coded diagram. The blue lines in the histograms of Fig. \ref{omlp} indicate the best fit to the right-hand side $N > N^{*}$ of them (more details regarding the probability distribution functions (PDF) are given at the end of this section). With increasing value of $b$ the the most probable number $(N^{*})$ of iterations is increased from 7 when $b = 0.1$ to 24 when $b = 100$.

\subsection{Case II: High angular velocity}
\label{ss2}

We continue our exploration with the case where the angular velocity has a high value, that is when $\omega = 5$. The Newton-Raphson basins of attraction for four values of the mass parameter $b$ are presented in Fig. \ref{omh}(a-d). Looking at panels (a)-(d) of Fig. \ref{omh} we may identify the following phenomena which occur as the value of the mass parameter increases: (i) the area of the basins of attraction corresponding to the collinear points $L_2$ and $L_3$ is constantly reduced and for $b = 100$ it almost disappears; (ii) the structures of the basins of attraction corresponding to the triangular points $L_5$ and $L_6$ become wider, however at extremely high values of $b$ they do not merge, as it was seen in the previous case; (iii) the fractality at the basin boundaries of equilibrium points $L_1$, $L_4$, $L_5$, and $L_6$ increases.

At this point is should be emphasized that by looking at Fig. \ref{omh} one may wrongly conclude that when the value of the angular velocity is high the area of the basins of convergence of collinear points $L_2$ and $L_3$ is finite. However this assumption is entirely wrong. Our numerical calculations indicate that lonely points corresponding to these two libration points randomly exist mainly at the boundaries of the attracting domains of the other equilibrium points. Therefore the area of all the basins of attraction is infinite also in this case.

In Fig. \ref{omhn}(a-d) we illustrate the distribution of the corresponding number $(N)$ of iterations required for obtaining the desired accuracy. The corresponding probability distribution of iterations is given in Fig. \ref{omhp}(a-d). The the most probable number $(N^{*})$ of iteration starts at 8 for $b = 0.1$ and then it increases up to $b = 100$, where the highest value, $N^{*} = 20$ has been observed.

Before closing this numerical investigation we would like to shed some light to the probability distributions of iterations presented in Figs. \ref{omlp} and \ref{omhp}. In particular, it would be very interesting to try to obtain the best fit of the tails\footnote{By the term ``tails" of the distributions we refer to the right-hand side of the histograms, that is, for $N > N^{*}$.} of the distributions. For finding the best fit of the tails we tried several single types of distributions (Laplace, Maxwell-Boltzmann, Rayleigh, Pascal, Poisson, etc). Our calculations strongly indicate that in the vast majority of the cases the Laplace distribution is the best fit to our data.

The probability density function (PDF) of the Laplace distribution is given by
\begin{equation}
P(N | l,d) = \frac{1}{2d}
 \begin{cases}
      \exp\left(- \frac{l - N}{d} \right), & \text{if } N < l \\
      \exp\left(- \frac{N - l}{d} \right), & \text{if } N \geq l
 \end{cases},
\label{pdf}
\end{equation}
where $l$ is the location parameter, while $d > 0$, is the diversity. In our case we are interested only for the $x \geq l$ part of the distribution function.

In Table \ref{table1} we present the values of the location parameter $l$ and the diversity $d$, as they have obtained through the best fit, for all cases discussed in Figs. \ref{omlp} and \ref{omhp}. One may observe that for most of the cases the location parameter $l$ is very close to the most probable number $N^{*}$ of iterations, while in some cases these two quantities coincide.

Finally, we would like to emphasize that the Laplace distribution is only a first good approximation to our data. Additional numerical calculations indicate that if we use a mixture of several types of distributions, instead of a single type of distribution (i.e., the Laplace distribution), the fit is much better. However we feel that this task is out of the scope and the spirit of this paper and therefore we did not pursue it.

\begin{table}[!ht]
\begin{center}
   \caption{The values of the location parameter $l$ and the diversity $d$, related to the most probable number $N^{*}$ of iterations, for all the studied cases shown in Figs. \ref{omlp} and \ref{omhp}.}
   \label{table1}
   \setlength{\tabcolsep}{10pt}
   \begin{tabular}{@{}lrrrr}
      \hline
      Figure & $b$ & $N^{*}$ & $l$ & $d$ \\
      \hline
      \ref{omlp}a & 0.1 &  7 & $N^{*} +  6$ &  5.88358415 \\
      \ref{omlp}b &   1 &  8 & $N^{*} +  5$ &  6.10745048 \\
      \ref{omlp}c &   5 & 11 & $N^{*} + 10$ &  8.06824641 \\
      \ref{omlp}d &  10 & 22 & $N^{*}$      &  8.34333014 \\
      \ref{omlp}e &  50 & 24 & $N^{*} +  2$ &  9.45499095 \\
      \ref{omlp}f & 100 & 24 & $N^{*} +  4$ & 10.05573690 \\
      \hline
      \ref{omhp}a & 0.1 &  8 & $N^{*} + 2$  &  3.64874464 \\
      \ref{omhp}b &   1 & 13 & $N^{*} + 2$  &  5.02970652 \\
      \ref{omhp}c &   5 & 16 & $N^{*} + 3$  &  6.82627710 \\
      \ref{omhp}d & 100 & 20 & $N^{*} + 5$  &  9.79103435 \\
      \hline
   \end{tabular}
\end{center}
\end{table}

\section{Discussion and conclusions}
\label{disc}

The scope of this research paper was to numerically determine the basins of convergence associated with the equilibrium points. In the LRFBP the position and the type of the libration points strongly depends on the values of the mass parameter $b$ and the angular velocity $\omega$. With the help of the multivariate version of the Newton-Raphson iterative scheme we managed to unveil the extraordinary and magnificent structures of the basins of attraction corresponding to the equilibrium points of the dynamical system. These basins play an important role as they describe how each point on the configuration $(x,y)$ plane is attracted by the libration points which act as attractors. Our numerical exploration revealed how the position of the equilibrium points and of course the structure of the attracting areas are influenced by the mass parameter $b$ and the angular velocity $\omega$. Furthermore, we related the several basins of attraction with the corresponding distribution of the required number of iterations. To our knowledge, this is the first time that such a thorough and systematic numerical investigation, regarding the basins of attraction, takes place in the LRFBP and this is exactly the novelty as well as the importance of the current work.

The main results of our numerical exploration are the following:
\begin{enumerate}
  \item Our calculations strongly suggest that all the initial conditions of the configuration $(x,y)$ plane converge, sooner or later, to one of the six attractors. In other words, we did not encounter any non-converging points.
  \item In all examined cases the area of the basins of convergence corresponding to all equilibrium points is infinite.
  \item The iterative method was found to converge very fast $(0 \leq N < 15)$ for initial conditions around each equilibrium point, fast $(15 \leq N < 25)$ and slow $(25 \leq N < 50)$ for initial conditions that complement the central regions of the very fast convergence, and very slow $(N \geq 50)$ for initial conditions of dispersed points lying either in the vicinity of the basin boundaries, or between the dense regions of the equilibrium points.
  \item In general terms we concluded that the average value of required iterations $(N^{*})$ for obtaining the desired accuracy increases with increasing value of the mass parameter $b$. In almost all cases, the Newton-Raphson method, for more than 95\% of the initial conditions, requires less than 70 iterations to converge to one of the available attractors.
  \item Our tests indicate that our numerical data, corresponding to the histograms with the probability distributions of the required iterations, are best fitted by the Laplace probability distribution function (PDF). Only the cases just before the two critical values of the mass parameter (which have long tails) cannot be fitted well by a Laplace PDF.
\end{enumerate}

A double precision code, written in standard \verb!FORTRAN 77! \cite{PTVF92}, has been deployed for performing all the required numerical calculations regarding the basins of convergence. For the graphical illustration of the paper, we used the latest version 11.0 of Mathematica$^{\circledR}$ \cite{W03}. For the classification of each set of the initial conditions on the several types of two-dimensional planes, we needed about 5 minutes of CPU time using a Quad-Core i7 2.4 GHz PC, depending of course on the required number of iterations. When an initial condition had converged to one of the attractors with the predefined accuracy the iterative procedure was effectively ended and proceeded to the next available initial condition.

Judging by the novel results revealed through the detailed and systematic numerical exploration we believe that we successfully completed our computational task. We hope that our investigation and the corresponding outcomes to be useful in the field of attracting domains in the LRFBP. Taking into account that the current analysis was encouraging it is our future plans to try and use other types of iterative formulae (of higher order with respect to the classical Newton-Raphson method) and compare the similarities as well as the differences of the structures of the basins of attraction. Furthermore, it would be very interesting to determine how the structure of the attracting domains of the LRFBP changes when additional parameters (i.e., the oblateness or the radiation pressure) are taken into account.

\end{document}